# Waveguide-integrated single-crystalline GaP resonators on diamond

Nicole Thomas,[1,*] Russell J. Barbour,[2] Yuncheng Song,[3] Minjoo Larry Lee,[3] and Kai-Mei C. Fu[1,2]

[1]*Department of Electrical Engineering, University of Washington, 185 Stevens Way, Seattle, WA 98195, USA*
[2]*Department of Physics, University of Washington, 3910 15th Ave. NE, Seattle, WA 98195, USA*
[3]*Department of Electrical Engineering, Yale University, 15 Prospect St., New Haven, CT 06511, USA*
[*]nkthomas@uw.edu

**Abstract:** Large-scale entanglement of nitrogen-vacancy (NV) centers in diamond will require integration of NV centers with optical networks. Toward this goal, we present the fabrication of single-crystalline gallium phosphide (GaP) resonator-waveguide coupled structures on diamond. We demonstrate coupling between 1 µm diameter GaP disk resonators and waveguides with a loaded Q factor of 3,800, and evaluate their potential for efficient photon collection if integrated with single photon emitters. This work opens a path toward scalable NV entanglement in the hybrid GaP/diamond platform, with the potential to integrate on-chip photon collection, switching, and detection for applications in quantum information processing.

## 1. Introduction

Entanglement between numerous qubits provides an essential resource for quantum information applications such as measurement-based quantum computing [1-3] and quantum communication [4-6]. The negatively charged nitrogen-vacancy ($NV^-$) center in diamond is a promising system for quantum entanglement generation due to its long spin coherence time and spin-coupled optical transitions [7-10]. Integrating numerous $NV^-$ centers in diamond with solid state photonic networks could enable scalable on-chip entanglement generation between $NV^-$ electron spins [11]. Here, we demonstrate waveguide-integrated single-crystalline gallium phosphide (GaP) resonators on diamond as a building block for such quantum information processing (QIP) networks.

Entanglement generation in photonic QIP networks will require efficient on-chip photon collection and routing, photon interference, and single photon detection modules. Efficient photon collection and routing can be achieved through microcavity structures coupled to bus waveguides and optical modulators. In a GaP (n=3.31 at 637 nm) waveguiding layer on diamond (n=2.4), the evanescent field from the guided mode can overlap with near-surface NV⁻ centers. GaP is a linear electro-optic material ($\chi^{(2)}$ = -0.97 pm/V [12]), opening a path toward the realization of active optical switches for photon routing [13,14]. The realization of active devices is an outstanding challenge for all-diamond photonic platforms [15, 16]. Finally, on-chip superconducting single photon detectors are compatible with III-V semiconductor waveguides [17].

Coupling of the optical emission of near-surface NV⁻ defects in diamond to isolated GaP optical resonators and Purcell enhancement in these structures has previously been demonstrated [18,19]. However, the fabrication approach utilized does not allow for the deterministic placement of those resonators on the substrate, and hence cannot be employed for the fabrication of waveguide-integrated cavity structures. Ideally, a large GaP layer bonded with diamond would be used for device fabrication. One possible approach, direct growth of GaP on diamond substrates, currently results in polycrystalline GaP layers exhibiting high optical losses [20]. As an alternative, here we present the scalable transfer of low loss single-crystalline GaP sheets onto diamond to provide a substrate suitable for QIP optical networks. We demonstrate coupled GaP whispering gallery mode resonator-waveguide structures on diamond, integrated with grating couplers for free-space coupling. The devices are evaluated in terms of their potential photon collection efficiency for a near-surface NV⁻ center. We estimate a collection efficiency exceeding 20% for current devices, exceeding free-space coupling by over an order of magnitude [21]. If integrated with single NV⁻ centers, the devices provide the photon collection unit for QIP networks used to generate NV⁻ electron spin entanglement.

## 2. Figure of merit for single emitter photon collection efficiency

The success probability for two spin entanglement generation scales as $\eta^2$, where $\eta$ is the overall detection efficiency for a single zero-phonon-line (ZPL) photon emitted from the NV⁻ center [3]. The primary function of a waveguide-coupled resonator is the efficient collection of this ZPL emission into a useful waveguide mode. The probability for a coupled NV⁻ center to decay via a ZPL photon into a waveguide mode is given by

$$\eta_{NV-WG} = \eta_{ZPL} \times \beta_{ZPL} \times \eta_{cav-WG} \tag{1}$$

in which $\eta_{ZPL}$ is the fraction of total NV⁻ emission into the ZPL, $\beta_{ZPL}$ is the fraction of the ZPL emission coupled into the cavity, and $\eta_{cav-WG}$ is the fraction of the cavity field coupled into the waveguide. $\eta_{ZPL}$ is defined as

$$\eta_{ZPL} = \frac{\gamma_{ZPL}(F_{ZPL}+1)}{\gamma_{PSB}+\gamma_{ZPL}(F_{ZPL}+1)}, \tag{2}$$

in which $\gamma_{ZPL}$ and $\gamma_{PSB}$ are the spontaneous emission rates into the NV⁻ ZPL and phonon sideband (PSB), respectively. The ZPL emission makes up only a small fraction of the total spontaneous emission rate $\gamma_{NV}$: $\gamma_{ZPL}/\gamma_{NV} \sim 3\%$ [11,22,23]. However, if an NV⁻ is placed within the field of a hybrid GaP/diamond cavity, its emission rate into the ZPL can be enhanced by a Purcell factor $F_{ZPL}$ [19,24]. Under weak coupling, the enhancement is given by

$$F_{ZPL} = \frac{3}{4\pi^2}\left(\frac{\lambda_{ZPL}}{n_{GaP}}\right)^3 \frac{n_{GaP}}{n_{diamond}} \frac{Q_l}{V} \left(\frac{|\hat{\mu}\cdot E_{NV}|}{|E_{\max}|}\right)^2 \tag{3}$$

with $\lambda_{ZPL}$ the ZPL wavelength, $n_{GaP}$ and $n_{diamond}$ the refractive indeces of GaP and diamond, respectively, $Q_l$ the loaded cavity quality factor, $V$ the whispering gallery mode volume, $\hat{\mu}$ the NV⁻ dipole moment unit vector, and $E_{NV}$ and $E_{max}$ the local electric field at the defect and the maximum field in the cavity, respectively. The Purcell factor further determines the fraction of the ZPL emission coupled into a cavity:

$$\beta_{ZPL} = \frac{F_{ZPL}}{F_{ZPL}+1}. \tag{4}$$

The fraction of the cavity field coupled to an adjacent waveguide,

$$\eta_{cav-WG} = \frac{\gamma_{waveguide}}{\gamma_{total}} = \frac{Q_c}{Q_l} = \frac{Q_i}{Q_i+Q_c}, \tag{5}$$

is a function of the intrinsic quality factor $Q_i$, the coupling quality factor $Q_c$, and the loaded quality factor $Q_l$ of the waveguide-coupled cavity, assuming no scattering losses in the system. Thus, the total fraction of the NV$^-$ emission coupled into the bus waveguide (with the field travelling in both directions along the waveguide length) can be expressed as

$$\eta_{NV-WG} = \frac{\gamma_{ZPL}}{\gamma_{PSB}+\gamma_{ZPL}(F_{ZPL}+1)} \times F_{ZPL} \times \frac{Q_i}{Q_c+Q_i}. \tag{6}$$

Due to the strong PSB emission in the NV$^-$ system, the Purcell enhanced emission is typically << total NV$^-$ emission. In this limit, it can be shown that the ZPL emission is most efficiently coupled into the waveguide with the cavity close to critical coupling, $Q_i=Q_c$. We will use the derived figure of merit $\eta_{NV-WG}$ to evaluate the performance of our GaP waveguide-coupled resonators, and to assess the robustness of device performance to fabrication tolerances.

## 3. Transfer of single-crystalline GaP onto diamond

Transfer of submicrometer thick GaP sheets onto the mechanical grade diamond substrate for subsequent device fabrication was achieved using an epitaxial lift-off procedure modified from the one developed by Yablonovitch et al. for GaAs [25,26]. The GaP sample is prepared via molecular beam epitaxy of a GaP buffer layer, 800 nm of sacrificial $Al_{0.85}Ga_{0.15}P$, and the 200 nm GaP device layer, respectively, on a single-crystalline GaP substrate. The GaP root-mean-square surface roughness is ~ 0.81 nm, as measured by atomic force microscopy.

Figure 1(a) displays a schematic process flow for the release and transfer of the epitaxial GaP layer onto diamond. A 1 μm thick photoresist layer is spun onto the sample and patterned via photolithography. The pattern defines the outline of the GaP area to release as well as etch vias within this area. The etch vias are necessary for an efficient mass transfer of etchants and etch products during the release. The epilayer substrate is then etched in an inductively coupled plasma reactive ion etch (ICP-RIE) system using $Cl_2$/Ar etch chemistry to define the GaP transfer sheet. Release is accomplished by the removal of the sacrificial AlGaP layer with a diluted HF wet etch (1:100 49% HF : DI $H_2O$). The photoresist layer now acts as a mechanical support for the released layer. The GaP sheet is then transferred onto the diamond substrate to which it binds via van der Waals forces. Removal of the photoresist layer in hot 1165 resist remover concludes the transfer process. An optical micrograph of a transferred GaP sheet on a diamond substrate is shown in Fig. 1(b).

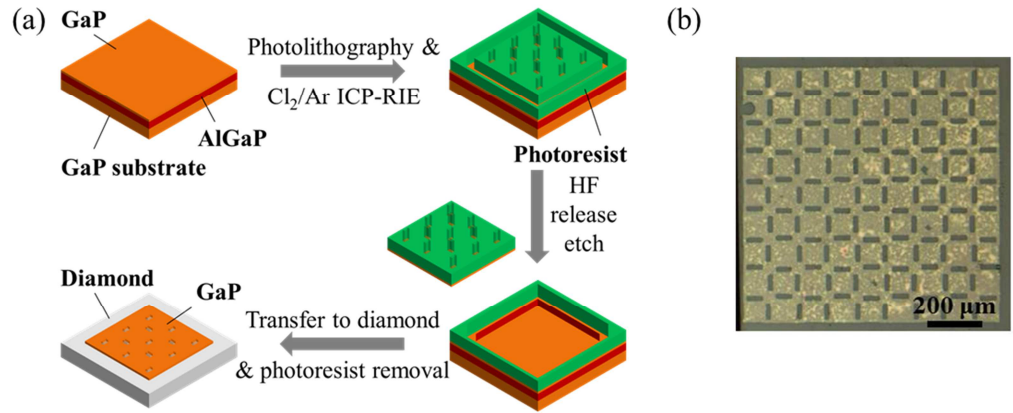

Fig. 1. Transfer of single-crystalline GaP sheets onto diamond. (a) Schematic process flow. (b) Optical micrograph of mm$^2$-sized transferred GaP sheet on diamond.

## 4. GaP resonator-waveguide structures on diamond

### 4.1 Fabrication of waveguide-coupled GaP resonators on diamond

Transfer of the 1 mm$^2$ sized GaP sheet onto the diamond substrate allows for the fabrication of coupled GaP resonators, waveguides and gratings for free-space coupling directly on diamond. We fabricate ring and disk resonators with diameters of 1 μm, 2 μm, and 5 μm and characteristic waveguide widths and resonator-waveguide gaps of 100 nm and 80 nm, respectively.

The devices are patterned in negative hydrogen silsesquioxane (HSQ) resist with a 10 nm SiO$_2$ adhesion layer using a 100 kV electron beam lithography system. The GaP layer is subsequently etched by Cl$_2$/Ar ICP-RIE. HSQ resist remains on the devices after processing. After initial measurements on the devices, we etch the diamond substrate using an O$_2$ etch chemistry to further confine the optical modes in the microcavity structures, and thereby enhance their intrinsic $Q_i$ factor [18,19]. Figure 2(a) shows a scanning electron micrograph (SEM) of an array of GaP devices on diamond.

### 4.2 Broadband transmission measurement setup

We test coupling between waveguides and resonators by measuring the transmission spectrum of a broadband light emitting diode (LED) source (640 nm) at room temperature. The structures are excited with the light coupled into the waveguides using out-of-plane grating couplers (Fig. 2(b)): the LED output is fiber-coupled and focused on the grating structure; the signal transmitted through the waveguide is then collected through the same lens using a second (output) grating coupler. Dips in the transmission spectrum are observed due to whispering gallery resonances of the disk or ring resonators.

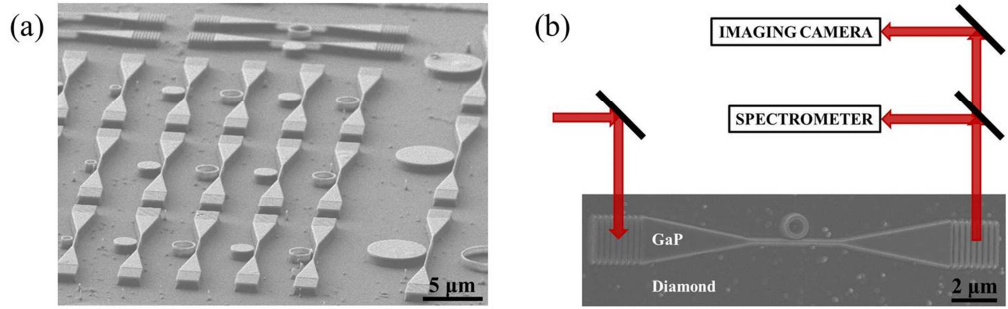

Fig. 2. GaP waveguide-integrated resonators with out-of-plane grating couplers on diamond. (a) Scanning electron micrograph of GaP resonator array on the etched diamond substrate. We fabricate ring and disk resonators with 1 μm, 2 μm, and 5 μm diameter, respectively. The waveguides are typically 100 nm wide, typical resonator-waveguide spacings are 80 nm. (b) Schematic of setup for resonator-waveguide coupling testing. We record the transmission spectrum from a broadband source used to excite the structure.

### 4.3 Device characteristics

Transmission measurements on the pristine diamond substrate reveal resonance dips in the spectra of 2 μm disk, and 5 μm disk and ring resonators only. As discussed below, this is theoretically expected due to low optical confinement in 1 μm diameter devices on the high-index diamond substrate. Figure 3(a) displays the broadband transmission spectrum recorded for a 2 μm disk resonator with sharp transmission dips at 618.5 nm, 636.4 nm, and 655.8 nm, due to whispering gallery mode resonances. Low-frequency oscillations caused by Fabry-Perot interferences along the length of the waveguide are also observed. The maximum loaded quality factor $Q_l$ extracted from the data is 3,700 at 636.4 nm, with a transmission dip of ~ 40%. Comparison of this data with cylindrical quasi 2-dimensional (2-D) FDTD [27] simulations indicates TE resonance modes with the measured free spectral range (Fig. 3(b)). The simulated intrinsic $Q_i$ is ~ 6,650 for the 640.2 nm TE resonance. This indicates that our fabricated optical circuit is under-coupled since the measured loaded $Q_l$ is higher than the maximum critically coupled $Q_l$ of ~ 3,325 expected for an ideal device.

FDTD simulations of 2 μm GaP devices with the diamond substrate etched 600 nm deep show significantly higher $Q_i$ factors (>$10^9$), due to higher confinement of the optical mode in the cavity (Fig. 3(c)) [18,19]. We further expect resonance modes to appear in 1 μm diameter GaP devices on diamond pedestals. We therefore etch 550 nm into the diamond substrate, and retest the same devices post-etch.

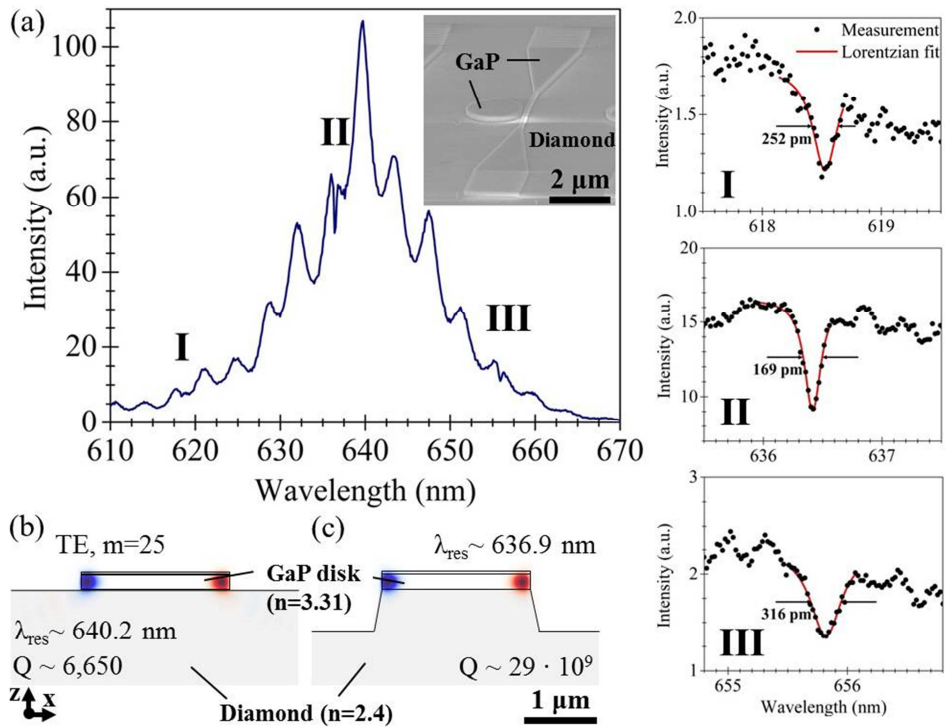

Fig. 3. Coupling between 2 μm diameter GaP disk and 100 nm GaP waveguide on diamond. (a) Transmission spectrum showing resonance dips at 618.5 nm (I), 636.4 nm (II), and 655.8 nm (III) with a maximum loaded $Q_l$ of 3,700. The broadband spectrum (left) is taken with a 300 lines/mm grating. The transmission spectra at the resonance wavelengths (right) is taken using a 1800 lines/mm grating. The inset shows an SEM image of a typical device. (b) Cylindrical quasi 2-D FDTD simulation of 2 μm diameter GaP disk on diamond showing a TE resonance at 640.2 nm with a $Q_i$ of 6,650, respectively. (c) FDTD simulation of 2 μm GaP disk on etched diamond pedestal. The TE resonance moved to 636.9 nm with a significantly increased $Q_i$ factor > 29 · $10^9$. Shown are the cross-sections of the resonators along y=0. All devices are simulated with a 40 nm $SiO_2$ top cladding.

Transmission spectra of structures on the etched diamond substrate show resonance dips for even the smallest size resonators. Figure 4(a) displays a SEM image of a 1 μm GaP disk resonator with a resonator-waveguide spacing of ~ 75 nm. The transmission spectrum in Fig.

4(b) shows a resonance at 643.5 nm with a loaded $Q_l$ of 3,800, with a dip of ~ 40%. The doublet structure of the dip may be caused by surface roughness and other imperfections, leading to non-degeneracy of the clockwise and counter-clockwise propagating modes in the resonator [28-30]. FDTD simulations of a pristine 1 μm GaP disk on the etched substrate show a $Q_i > 3·10^6$ for the TE resonance. In comparison, the simulated $Q_i$ for a 1 μm disk on a non-etched diamond substrate is ~ 200. The significantly smaller measured $Q_l$ points toward a combination of over-coupling between the fabricated structures and fabrication imperfections that result in rough sidewalls, scattering loss and a lower $Q_i$ [29,30]. 3-D FDTD simulations of a waveguide-integrated disk with 75 nm waveguide-resonator spacing indicate a coupling $Q_c$ of ~ 6,000. This suggests that our devices are indeed over-coupled, and have an estimated intrinsic $Q_i$ of ~ 10,000. Improvements in resist and etch processing are expected to yield higher quality devices: cavities fabricated from a comparable epilayer substrate using a resist reflow technique showed intrinsic $Q_i$ of > 280,000 for a 6 μm GaP resonator in air [31].

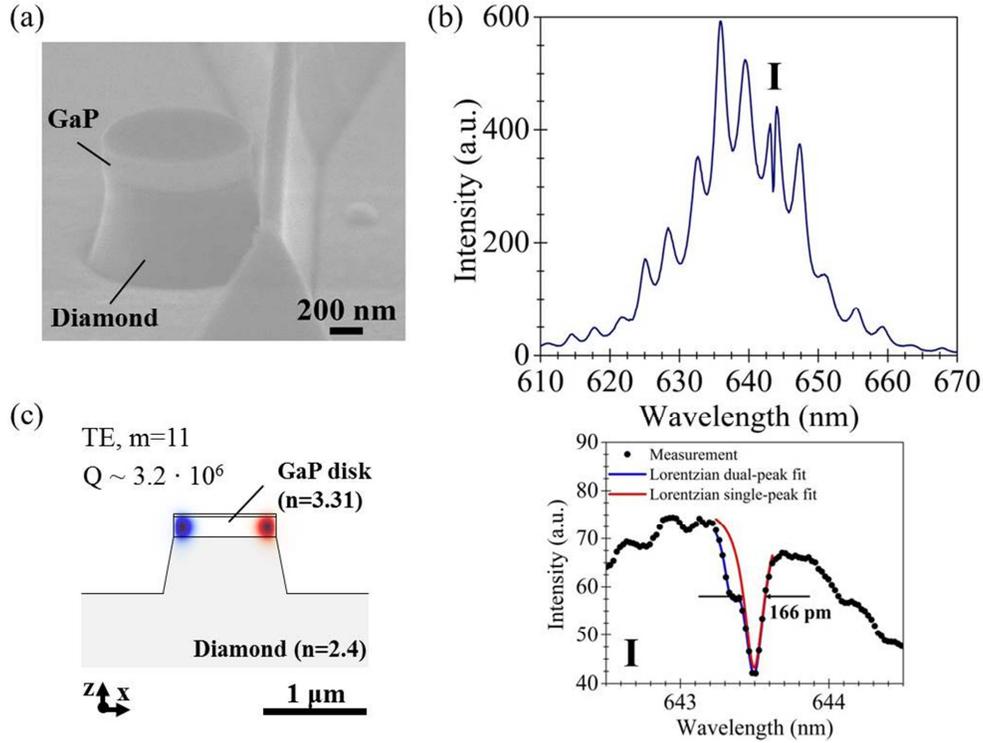

Fig. 4. Transmission spectrum for 1 μm diameter GaP disk resonator on etched diamond substrate. (a) SEM image of device with a waveguide-resonator separation of ~ 75 nm. (b) Broadband transmission spectrum showing a resonance dip at 643.5 nm with a $Q_l$ of 3,800. (c) Quasi 2-D FDTD simulation of a 1 μm GaP disk resonator on a diamond pedestal (cross-sectional view along y=0) with a quality factor $Q_i$ of 3.2·10$^6$.

## 5. Potential of waveguide-coupled disk resonators for integration with near-surface NV⁻

The successful implementation of hybrid GaP/diamond networks for QIP applications will require the effective enhancement of the NV⁻ emission rate and acceptable collection yield from the diamond substrate. As discussed in section 2, the Purcell factor is a crucial parameter for the efficient collection of photons from single defects in diamond. With our current 1 μm diameter devices exhibiting a $Q_l$ ~ 3,800 and an estimated $Q_i$ ~ 10,000, we expect a maximum Purcell enhancement $F_{ZPL}$ ~ 18 for the ZPL emission of a NV⁻ center located ~ 15 nm below the diamond surface [19]. We assumed optimal alignment of the NV⁻ dipole with the electric field in the 1 μm GaP disk resonator, and the NV⁻ location at an anti-node of the TE standing wave. Mode volume and the local field strength 15 nm below the diamond surface are extracted from 3-D FDTD simulations.

With an NV⁻ ZPL relative emission rate $\gamma_{ZPL}/\gamma_{NV} \sim 3\%$ [11,22,23], $F_{ZPL} \sim 18$, $Q_i \sim 10{,}000$, and $Q_c \sim 6{,}000$, we can estimate an overall collection efficiency in a bus waveguide of $\eta \sim 22\%$ in our current devices (Eq. 8). This number is only slightly enhanced under critical coupling conditions ($\eta \sim 26\%$). Indeed, only minor changes in $\eta$ can be observed for $Q_i \sim 10{,}000$ and $Q_c$ ranging between 5,000 ($\eta \sim 22\%$) and 15,000 ($\eta \sim 19\%$), i.e. a range in which the optical circuit is appreciably over- and under-coupled. Using the quality factors simulated in Fig. 5, we calculate an overall collection efficiency $\eta > 18\%$ for resonator-waveguide gaps between 65 nm and 105 nm, for both 100 nm and 120 nm wide GaP waveguides. This indicates a significant tolerance window for fabrication imperfections in both the waveguide width and the gap between the resonator and the waveguide, without detrimental effects on the overall NV⁻ ZPL collection efficiency.

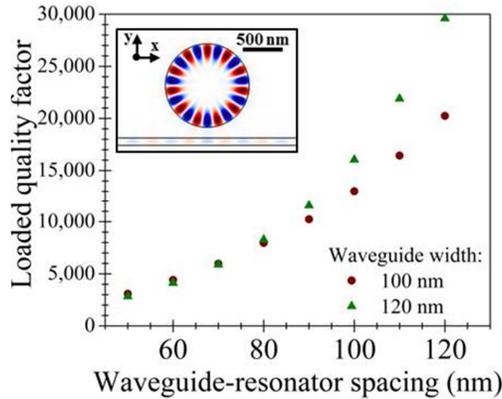

Fig. 5. 3D-FDTD simulation of loaded quality factor of a 1 μm diameter GaP disk on a diamond pedestal as function of the waveguide-resonator gap and waveguide width. The GaP devices with a diameter of 1 μm and height of 200 nm sits on a 550 nm deep etched diamond substrate. $Q_i$ of the simulated resonator is $> 3.2 \cdot 10^6$. The simulated loaded $Q_l$ are therefore mainly determined by the coupling $Q_c$. The inset shows phase-matched coupling between the GaP resonator and a 100 nm wide GaP waveguide with a gap of 100 nm, with the source located in the resonator.

## 6. Conclusion

We demonstrate for the first time waveguide-integrated GaP resonators on diamond. Realization of the hybrid devices is facilitated by the transfer of epitaxial, single-crystalline GaP sheets onto diamond. Loaded quality factors $Q_l$ of 3,700 for 2 μm diameter disk resonators on the pristine diamond substrate, and $Q_l$ of 3,800 for 1 μm GaP disk resonators on etched diamond pedestals were measured. We introduced a figure of merit for the total collection efficiency in a bus waveguide for photons emitted from a NV⁻ defect into a hybrid GaP/diamond resonator-waveguide system, and show that current devices are capable of NV⁻ photon collection efficiencies exceeding 20%. Moreover, device performance is found to be relatively robust to fabrication imperfections. This work marks a critical step toward an integrated platform for scalable quantum entanglement generation between NV⁻ centers for quantum information applications.

**Acknowledgment**

This material is based upon work supported by the National Science Foundation under Grant No. 1342902. N. Thomas acknowledges financial support from Intel Corp. Part of this work was conducted at the Washington Nanofabrication Facility, a member of the NSF National Nanotechnology Infrastructure Network. We are grateful to Richard Bojko for technical support for electron beam lithography.